\documentclass[conference, onecolumn]{IEEEtran}

\usepackage{xcolor}
\usepackage{tabularx}
\usepackage{booktabs}
\usepackage{graphicx}
\usepackage{algorithm2e}

\begin{document}
\title{Fully Automated Artery-Vein ratio and vascular tortuosity measurement in retinal fundus images}

\author{\IEEEauthorblockN{Aashis Khanal}
\IEEEauthorblockA{Department of Computer Science \\
Georgia State University, Atlanta, Georgia 30303\\
akhanal1@student.gsu.edu}

\IEEEauthorblockN{Rolando Estrada}
\IEEEauthorblockA{Department of Computer Science \\
Georgia State University, Atlanta, Georgia 30303\\
restrada1@student.gsu.edu}
}

\maketitle
\begin{abstract}
Accurate measurements of abnormalities like Artery-Vein ratio and tortuosity in fundus images is an actively researched task.
Most of the research seems to compute such features independently. However, in this work, we have devised a fully automated
technique to measure any vascular abnormalities. This paper is a follow-up paper on vessel topology estimation and extraction–we
use the extracted topology to perform A-V state of the art Artery-Vein classification, A-V ratio calculation, and vessel tortuosity
measurement, all fully automated. Existing techniques tend to only work on the partial region but we extract the complete vascular
structure. We have shown the usability of this topology by extracting two of the most important vascular features; Artery-Vein
ratio, and vessel tortuosity.
\end{abstract}
\IEEEpeerreviewmaketitle

\section{Introduction}
The research in automating diseases associated to the eye have been done extensively in the recent years. Especially we focus on 2D retinal fundus images, which are image taken of the back of the retina by a non-invasive device called funduscope. The main goal of this procedure is to identify abnormalities in the eye that could lead to very serious optical diseases like Diabetic Retinopathy \cite{retinal_imaging_and_analysis-5660089}, Glaucoma\cite{Glaucoma_fact_Schuster2020-px} and could even lead to blindness. As per Centers for Disease Control and Prevention(CDC), USA, 1 in 5 people have some form of diabetes, the cause of Diabetic Retinopathy. Early detection of such diseases can save huge economic burden \cite{economic_analysis_pmid10868871}. Similarly Glaucoma is one of the top-3 leading cause of blindness with a huge economic burden\cite{Glaucoma_fact_Schuster2020-px}. The research on automated diagnosis seemed to be divided into two main categories.

\begin{figure*}[!ht]
    \centering
    \includegraphics[width=0.99\textwidth]{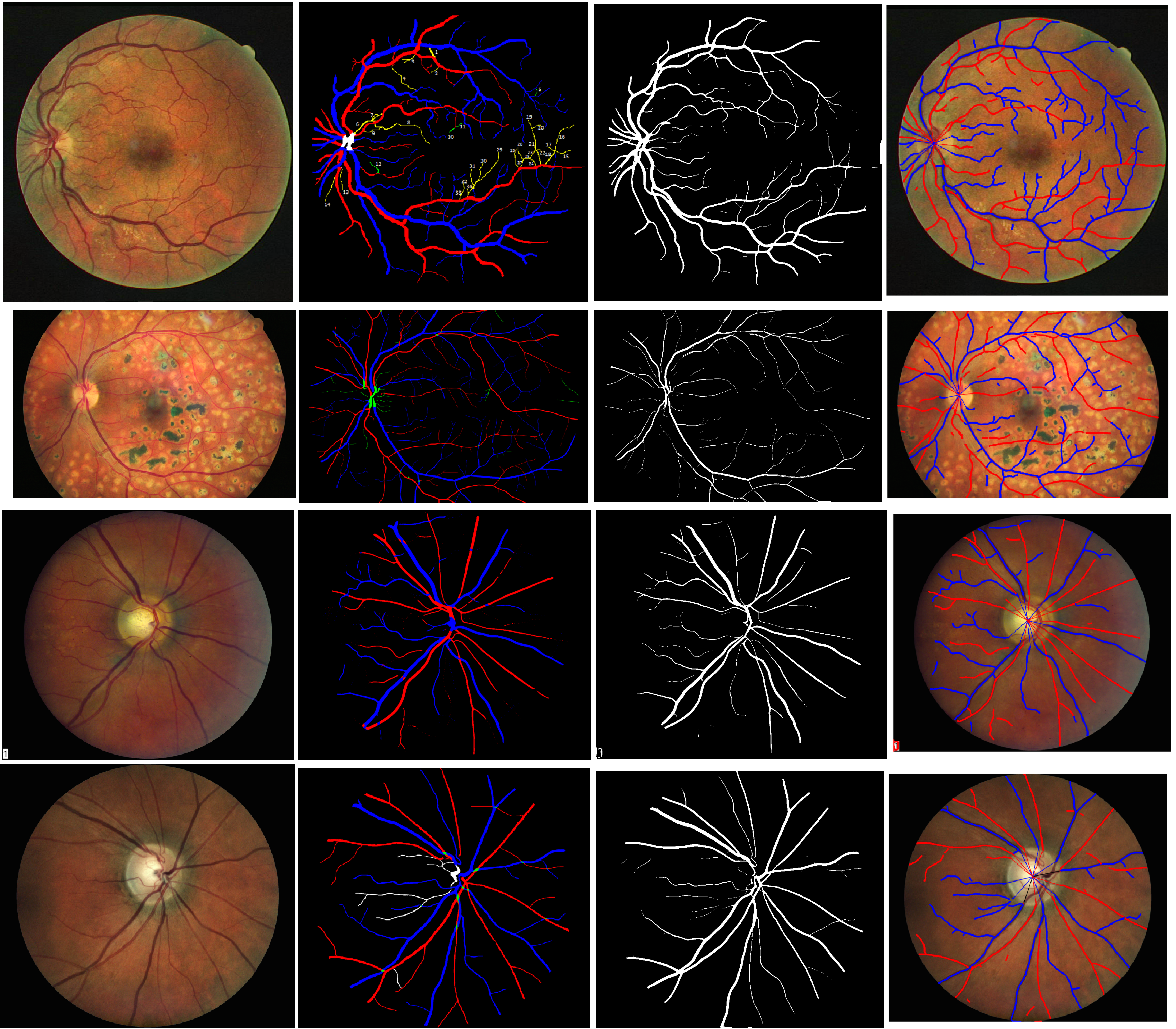}
    \caption{Datasets and results: DRIVE\cite{drive_dataset}, HRF\cite{HRF_dataset_Budai2013, hrf_AV_hemelings2019avs}, INSPIRE\cite{inspire_dataset-tang2011robust, inspire_av_dataset}, and LES\cite{LES_dataset} datasets from top to bottom. Original image, Artery-vein ground truth, vessel mask with a single threshold of $100(0-255)$, and final topology graph with Artery-Vein prediction.}
    \label{fig:DatasetsResults}
\end{figure*}

In \textbf{Direct diagnosis} approach, we usually feed the entire fundus image to a model like a Convolutional Neural Network(CNN), and ask for a likelihood for a number of diseases \cite{DL_DR_MacularEdma_Sahlsten2019, DR_AMD_pmid21666234, 10.1001/jama.2016.17216,six_disease_10.3389/fmed.2022.808402}. This process seem to yield good results in terms of measurements on the validation and test set, however the important factor of explain-ability seem to be absent. Ophthalmologist, and even patients tend to ask for a explainable diagnosis rather that a final conclusion for any medical condition. This method is notorious for needing a very large amount of labeled data as well, which is very time consuming especially in medical field. \textbf{Explained Feature based diagnosis} is another pathway is doing an explained diagnosis as an ophthalmologist would do, where we get a clear explanation on why a particular diagnosis was given for a patient. The existing work on this field is limited to either AV-segmentation, or AV-ratio, and tortuosity measurement independently. This paper is aimed to move this line of research to a significant step further by introducing a better way of extracting the complete vascular structure and calculating two of the most important abnormalities; AV-ratio, and vessel tortuosity. We strongly believe one can extract other vascular features with minimum effort from here as needed.

This work is a very natural extension of our previous work that extracts the complete vascular topology \cite{top_https://doi.org/10.48550/arxiv.2202.02382}. Additionally, we introduce a vessel caliber based smoothing mechanism to center-line the vessel path graph to remove local noise. This yields a topology graph as seen in figure \ref{fig:DatasetsResults} last column. The third column is the binarized vessel mask by using a fixed threshold of $100$. We can see that that a fixed threshold suffers from the precision-recall trade off--some low confident vessel are disconnected, if we use a very low threshold, chances are, it will pick up noise in some cases, where as in others it merges two vessels passing very closely as one. This poses a huge problem in measuring vessel features like width and tortuosity. However, we can see the vascular structure extracted using the topology extraction algorithm yields much more complete vasculature with the same likelihood map. For the same reason, we have yielded state of the art A-V classification results by training to predict the nodes of this graph with the help of some transfer learning (figure \ref{fig:DatasetsResults}, last column).

We have also calculated state-of-the art \textbf{A-V ratio} on the INSPIRE-AVR dataset \cite{inspire_av_dataset} as shown in figure \ref{fig:AVR-sample}. Which is calculated by extracting top six wide arteries and six wide veins within $1D$ to $1.5D$ from the disc center and taking the ratio. Since this dataset does not include optic disc mask, we use transfer learning to extract it. Similarly, we also calculate the tortuosity of each vessels within $1.5D-2.5D$($D$ being the diameter) from disc center as in figure \ref{fig:Tortuosity}.

\section{Related work}
Our main goal is extraction of vascular abnormalities thus we focus on tasks related as such. Mainly there are three tasks--vessel segmentation, artery-vein classification, topology extraction. We also need optic disk to calculate vascular features like artery-vein ratio, so it also naturally involves optic disk segmentation. The vessel segmentation and A-V segmentation is usually done as a pixel wise segmentation using convolutional neural networks \cite{uncertainty_aware_8759380,cnn_mst_8309054,multi_task_https://doi.org/10.48550/arxiv.2007.09337,Hu2021}. Others have done by extracting the topology and using it as a prior \cite{graphav1_6517259,7120990,top_aware_shin_lee_yun_lee_2020}. These technique rely on hard binarization of the likelihood map and begin the post processing like label propagation, and topology extraction. The figure \ref{fig:DatasetsResults}($3_{rd}$ column from left) shows that using such hard threshold posed a very complicated precision recall trade-off scenario. However, looking at the vasculature extracted from our topology estimation algorithm(last column from left), we can see the problem of such trade-off is non-existent. Since the A-V label prediction is dependent on how well we can extract the vessels, the error propagates to downstream tasks. In our approach we run A-V prediction of the topology nodes rather than pixels, making the network more aware of the existence of a vasculature.

There research in the feature calculation are very limited. The only with the quantified experiments is Niemeijer et al.\cite{inspire_av_dataset} proposed a method to calculate A-V ratio involving segmentation, and center line detection, however it is very limiting as this process cannot be used to extract other vascular features. However, being a complete vascular structure we can extract much more features. Comparatively, our method also achieves state of the art A-V segmentation with the baseline U-Net. Similarly, for vessel tortuosity, as well, there are only a handful of research works. Chen et al. \cite{vessel_seg_tort_app10144788} proposed a multi step curvature calculation framework, however the application is limited to selected segments of the vessels as opposed to the entire fundus image in our work. The most widely followed approach for vascular tortuosity measurement has been proposed by Grisan et al.\cite{grisan_turto_4359043} that splits a long vessel into smaller segments based on local curvature. This addresses the problem of inconsistent measures when the vessels are too large and the curve is naturally wide, but calculating arc-to-chord ratio yields a larger value. Based off of Grisan et al., Ramos et al. \cite{comp_assesment_tort_Ramos2019} proposed a global tortuosity measure which is a weighted sum of vessels parameterized by the distance to optic disk center, distance to fovea, A-V label. This poses a lot of point of failure as all the intermediate tasks like A-V classification, Optic DIsk/Fovea classification are still in research phase. Pur framework, however, allows the downstream use case to to pick either segments(vessel path between branch nodes), or select any caliber vessel, or work on just Arteries or Veins and many more.

\section{Method} 
\begin{figure*}[!ht]
    \centering
    \includegraphics[width=0.99\textwidth]{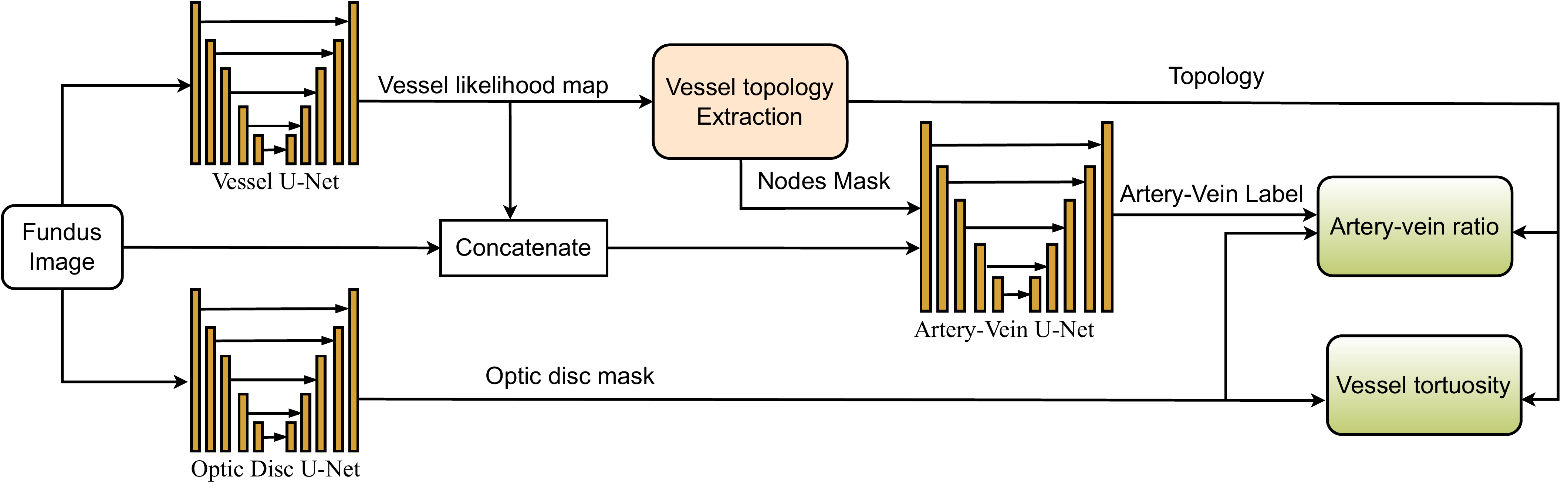}
    \caption{Artery-Vein ratio and vascular tortuosity calculation flow.}
    \label{fig:AVR_flow}
\end{figure*}

The entire pipeline is summarized in the figure \ref{fig:AVR_flow}. It starts with a fundus image that passes through a \textbf{Vessel-UNet} yielding a vessel likelihood map. We use the topology extraction method on the vessel pmap. The vessel likelihood is concatenated with the original image and passed through \textbf{Artery-vein UNet} to classify each nodes into either Artery or Vein. Similarly we also pass the image through \textbf{Optic Disc U-Net} to extract the optic disk as we need the diameter and the center for Artery-Vein ratio and tortuosity measurements.

\subsection{Topology extraction and smoothing} We use $5-fold$ cross validation to generate likelihood map for the entire datasets listed in table \ref{tab:datasets}. Then as in the original topology extraction paper, \cite{top_https://doi.org/10.48550/arxiv.2202.02382}. We use multiple thresholds on the likelihood map and skeletonize. Such skeletons are combined to form a union graph where each vessel pixel is a node, and two node share an edge if they are neighboring pixels in 8-connectivity. We the use a graph contraction mechanishm as explained in the original topology paper \cite{top_https://doi.org/10.48550/arxiv.2202.02382} to remove $>80\%$ fo the nodes and edges but still keep the overall vasculature intact. We weight each edge by two nodes'(or pixels') color, vessel likelihood, and vessel caliber, and run Dijkstra's shortest path algorithm. The path traced by Dijkstra algorithm might not follow the vessel's center-line path, thus we retrace the path of each segments(nodes path between two branch nodes). We weight each edge by the minimum distance of two pixels(represented as nodes in the graph) to the background $B_w$, which is normalized as in algorithm \ref{algo1:bw_norm}. This strategy makes traversing through pixels that have high likelihood of being a vessel less costly, yet being in the center of the vessel.

\begin{algorithm}
\SetAlgoLined
$T$ = set of threshold for likelihood map binarization as in \cite{top_https://doi.org/10.48550/arxiv.2202.02382} in increasing order. \\
$B_w$ = $0_{W \times H}$ (Same shape as the likelihood map)\\
\For{i, t in (\textbf{range}($len(T)$), $T$)}{
    \textbf{if} $i>5$:  $B_w$ +=  $log(1 + bwDist(seg_{t}) \times i^2)$ \\
    \textbf{else}: $B_w$ +=  $log(1 + bwDist(seg_{t}))$
}
\caption{Normalized minimum distance to background, where $bwDist$ is the distance transform, which is the distance to the closet background pixel in euclidean space.}
\label{algo1:bw_norm}
\end{algorithm}

\subsection{Optic Disk/Vessel extraction, and Artery-Vein classification with CNN} There are three CNN involved in our pipeline as shown in figure \ref{fig:AVR_flow}. We have used U-Net as the backbone for all of them \cite{Ronneberger2015UNetCN}. This architecture is widely popular in medical images for its performance superiority. It consist of multiple upscale-downscale of feature maps with skip connection on each levels. We used the \textbf{Vessel U-Net} to generate the vessel likelihood for each pixels. The topology extraction algorithm takes this likelihood map and the original image and outputs the topology where we represent the vasculature as nodes and edges\cite{top_https://doi.org/10.48550/arxiv.2202.02382}. Similarly, in parallel, we use \textbf{Optic Disk U-Net}, that only takes the original image, to extract the optic disk. Finally the \textbf{Artery-Vein U-Net} takes the original image, vessel likelihood map, and the nodes mask obtained from the topology and gives prediction to each of the nodes to either Artery or Vein. The traditional way of A-V classification is based on extracting the vessel mask and then classifying each of those pixels to Artery-Vein. Our technique first extracts the complete vasculature and does the classification. This ensures we capture all the vessels, as opposed to using a mask, where fainter vessel have very high change of being ignored. The results(table \ref{tab:literature-results}) have shown that, even with a simple base U-Net yields comparable to state-of-the art classification results. 

\subsubsection{Transfer learning for optic disk and vessel likelihood map} The INSPIRE-AVR\cite{inspire_av_dataset} dataset of which we want to calculate AV-ratio, and tortuosity, we don't have vessel ground truth, and optic disk mask available so we use transfer learning to extract those. We trained a single model using easytorch\cite{easytorch} framework on DRIONS\cite{drions_db_paper_dataset}, DRISHTI \cite{dristi_paper_dataset}, DRIVE\cite{drive_dataset,dcpa_2110.00512}, WIDE\cite{wide_dataset:6987362, dcpa_2110.00512}, CHASE-DB\cite{chasedb_dataset} and used that to generate optic disk mask for the INSPIRE-AVR\cite{inspire_av_dataset} dataset. Similarly, for vessel likelihood map of the INSPIRE dataset, we trained another model with DRIVE, STARE\cite{stare_dataset_845178}, WIDE, CHASEDB, HRF, IOSTAR\cite{iostar_7530915} and used the trained weights.

\subsubsection{Transfer learning for Artery-Vein classification on topology nodes}
As in the figure \ref{fig:AVR_flow}, the \textbf{Artery-Vein U-Net} takes the original fundus image, a vessel likelihood, and the nodes mask obtained from the topology. We train this network to only classify nodes pixel by only penalizing the nodes mask during training, making the network easier to learn the vessel connectivity and various constraint Artery and Veins follow. Instead of training this CNN from scratch, we first trained the HRF dataset from scratch, as it was the easiest and gave the best result on the test set, then used the weights on all other dataset and train from there. Our experiments have shown that it helped the network converge better as shown in table \ref{tab:literature-results}.

\subsection{Artery Vein ratio measurement}
We measure the A-V ratio by taking six largest arteries and six largest veins(can be less if six are not available) within $1D$ - $1.5D$ range as show in the figure \ref{fig:AVR-sample}. To incorporate the discrepancy due to branching factor, Knudtson et al \cite{crae_crve_doi:10.1076/ceyr.27.3.143.16049} have proposed an iterative method that computes the ratio as $\frac{CRAE(A)}{CRVE(V)}$, where $A$ and $V$ are list of width of arteries and veins in descending order, $CRAE$ being Central Retinal Artery Equivalent, and $CRVE$ as Central Retinal Vein Equivalent. The calculate the actual width of each pariticipating vessels, we first binarize the vessel likelihood map by a threshold($t=100$), and calculate the distance transform $bwDist$ as discussed in algo. \ref{algo1:bw_norm}. If the topology vessel path passes through the center-line of each vessels, we can use the node's corresponding value from $bwDist$ as the half width. We made sure that the topology vessel path passes through the center by the smoothing algorithm \ref{algo1:bw_norm}. Once we have all the widths, the iterative process of their calculation is detailed in algorithm \ref{algo2:CRAE_CRVE}.

\begin{algorithm}
\SetAlgoLined
$C = []$; \\
$W$ = list of widths in descending order with length $min(len(A), len(V), 6)$; \\
\While{$len(A)>1$}{
    $f$ = $W.pop(0)$; // pop the first element \\
    $l$ = $W.pop(-1)$; // pop the last element \\
    $C.append(p \cdot \sqrt{f^2 + l^2})$; \\
    \If{$len(W)<=1$}{
        $W = W + C$; \\
        $sortDescending(A)$; \\
        $C = []$;
    } 
}
\caption{$CRAE$ and $CRVE$ calculation routine. The branching cefficient provided by Knudtson et al. \cite{crae_crve_doi:10.1076/ceyr.27.3.143.16049} is $p=0.88$ for arteries, and $p=0.95$ for veins. Once the routine completes, we will get a single value for arteries and, a single values for veins which we use to calculate the A-V ratio.}
\label{algo2:CRAE_CRVE}
\end{algorithm}

\subsubsection{Extraction of $top-6$ wide arteries and veins from the topology graph} 
\begin{figure*}[!ht]
    \centering
    \includegraphics[width=0.99\textwidth]{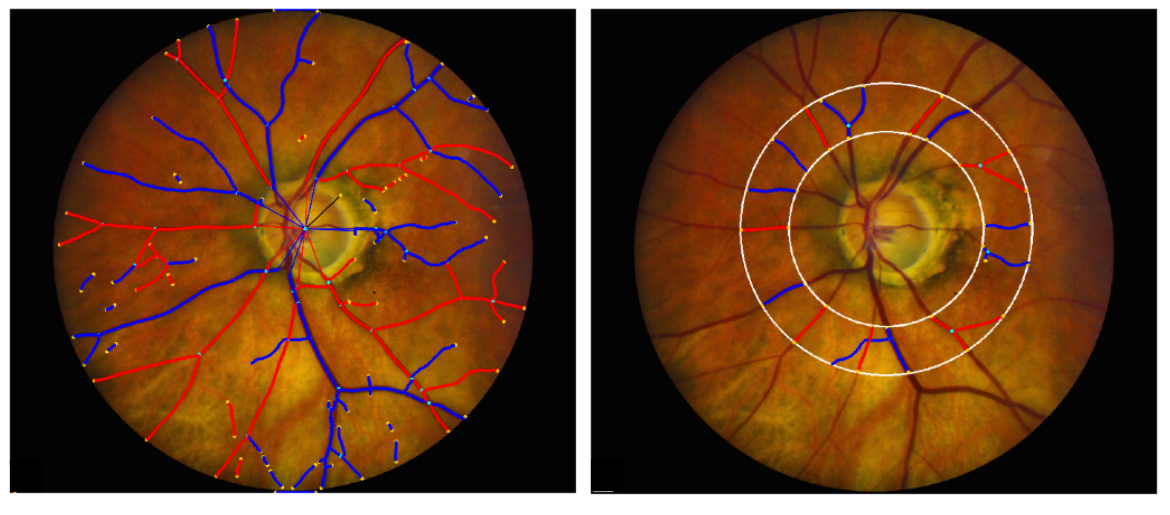}
    \caption{AVR Sample: Left image is the extracted topology graph with Artery-Vein label. Right is the region of $1 \times D$ to $1.5 \times D$ from the optic disc center, where $D$ being diameter of the optic disc. We take six most wide Arteries and Veins and take the ratio.}
    \label{fig:AVR-sample}
\end{figure*}

The figure \ref{fig:AVR-sample}, right). We remove the connected components that do not span both inner and outer circle(in white in fig. \ref{fig:AVR-sample}). We the use the route algorithm from the original topology paper \cite{top_https://doi.org/10.48550/arxiv.2202.02382}from the end nodes touching the inner circle to the end nodes touching the outer circle. We then remove the visited vessels and repeat the process in reverse order(from end nodes touching outer circle to inner). To calculate the actual width of each vessels, we take the $bwDist$ of nodes on $0.25, 0.5, 0.75$ length, and take the average. We group the vessel width by Artery and Vein based on A-V labels provided by our \textbf{Artery-Vein U-Net}--we simply label a vessel as an artery if the average likelihood of being artery, as given by the CNN, is higher than that of the veins, and vice versa. Once we have all the widths, we simply use the iterative algorithm \ref{algo2:CRAE_CRVE} to calculate the AV-ratio.

\subsection{Tortuosity measurement} To measure vessel tortuosity we only consider vessel in range $1D=2.5D$ from the optic Disk center as show in the figure \ref{fig:Tortuosity}. We extract the sub-graph within this region, remove branch vessels that are very small($<l$  nodes, where $l=10$). We also skip segments(nodes between two branch nodes), that are less than $2*l$. Now we have vessel path that span long distance but ignore ambiguous regions like branch points. To actually measure the tortuosity, we use the standard technique by Grisan et al. \cite{grisan_turto_4359043}, given by equation \ref{eq:grisen-torto}. The work also consist of method to detect such local curvature points and we simply used it out of the box. 

\begin{equation}
    \label{eq:grisen-torto}
    T_g = \frac{n-1}{L_c} \sum_{i=1}^{n} \bigg[  \frac{L_{c}^{i}}{L_{x}^{i}} - 1 \bigg]
\end{equation}, where $n$ is number of sub-segments, $L_c$ is chord length of the full segment, $L_{c}^{i}$ is the arc length and $L_{x}^{i}$ is the chord length.

\section{Experiments and results}
We have used four well known datasets in this work as described in table \ref{tab:datasets}. As we can see the dataset varies on resolution a lot, we decided to resize all images to maximum dimension of $896 \times 896$ maintaining the aspect ratio. This also shows the superiority of our technique in topology extraction when image are extremely re-scaled upward(DRIVE dataset), and downward(HRF dataset). The scores reported are of $5-fold$ cross validation on all datasets.  
\subsection{Neural Networks training}
We performed all the training in our intel server with 2 2080 ti NVIDIA Graphics card, 48 logical cores, and 256 GB Memory. We trained all the experiments with a patience of 50 epochs, where we stop the training if the $F_1$ score on validation set does not improve in last 50 epochs. We used stochastic class weights as mentioned in the paper \cite{deep_dyn_10.3389/fcomp.2020.00035} while training. We feed the neural networks $388 \times 388$ image patch, and use sliding window to feed the entire image. We use patch overlap of 250 pixels on both dimensions. We use Adam optimizer with learning rate of 0.001. 

\begin{table}[!ht]
\caption{Datasets used and their details}
\label{tab:datasets}
\centering
\begin{tabular}{lccp{5cm}}
 \textbf{Dataset} &  \textbf{Total images} & \textbf{Image resolution} & \textbf{Comments}  \\
 \toprule
 DRIVE \cite{drive_dataset} & 40 & 565 $\times$ 584 & Standard dataset used in almost all liteature.  \\
 HRF \cite{HRF_dataset_Budai2013,hrf_AV_hemelings2019avs} & 45 & 3504 $\times$ 2336 &  Ultra HD images. \\
 INSPIRE-AVR \cite{inspire_dataset-tang2011robust,inspire_av_dataset} & 40 & 2392 $\times$ 2048 & \\
 LES \cite{LES_dataset} & 22 & 1620 $\times$ 1444 &   \\
\bottomrule
\end{tabular}
\end{table}

\begin{table*}[ht!]
\caption{A/V segmentation result of existing techniques}
\label{tab:literature-results}
\setlength{\tabcolsep}{4pt}
\begin{center}
\begin{small}
    \begin{tabular}{rcccccccccccc}
        \toprule
         & \multicolumn{3}{c}{DRIVE} & \multicolumn{3}{c}{INSPIRE-AV} & \multicolumn{3}{c}{HRF} & \multicolumn{3}{c}{LES} \\
        \midrule 
        \textbf{Method} & \textbf{BACC} & \textbf{SEN} & \textbf{SPE} & \textbf{BACC} & \textbf{SEN} & \textbf{SPE} & \textbf{BACC} & \textbf{SEN} & \textbf{SPE} & \textbf{BACC} & \textbf{SEN} & \textbf{SPE}\\
        \midrule 
        
        UA-AV\cite{uncertainty_aware_8759380} & 0.895 & 0.890 & 0.900 & 0.800 & 0.820 & 0.780 & - & - & - & 0.865 & 0.880 & 0.850 \\
        
        Graph based\cite{graphav1_6517259} & 0.870 & 0.900 & 0.840 & 0.885 & 0.910 & 0.860 & - & - & -& - & - & - \\
        
        Via topology\cite{7120990} & 0.935 & 0.930 & 0.941  & 0.909 & 0.915 & 0.902 & - & - & -& - & - & - \\
        
        Score prop.\cite{cnn_mst_8309054} & 0.927 & 0.923 & 0.931 & - & - & - & - & - & -& - & - & -\\ 
        
        Multi task\cite{multi_task_https://doi.org/10.48550/arxiv.2007.09337} & 0.944 & 0.934 & 0.955 & 0.918 & 0.924 & 0.913 & - & - & -& - & - & - \\
        
        Vessel-constraint\cite{Hu2021} & 0.955 & 0.0.936 & 0.974 & - & - & - & 0.964 & 0.958 & 0.970 & 0.944 & 0.942 & 0.946 \\
        \midrule
        
        \textbf{CNN*} & 0.926 & 0.936 & 0.917 & 0.903 & 0.901 & 0.906 & 0.946 & 0.954 & 0.938 & 0.889 & 0.907 & 0.872 \\
        
        \textbf{CNN + Prop*} & 0.934 & 0.943 & 0.925 & 0.924 & 0.917 & 0.931 & 0.951 & 0.955 & 0.948 & 0.904 & 0.915 & 0.894 \\
        \midrule
        \multicolumn{12}{l}{\small * The scores reported are based on vessel centerline pixels, not the whole vessel mask.} \\
        
    \end{tabular}
\end{small}
\end{center}
\end{table*}

\subsection{Artery-Vein classification on topology}
Table \ref{tab:literature-results} shows the comparison between A-V classification with vessel mask vs nodes. As we can see using a vanilla U-Net to classify nodes yields very comparable results. One thing to note is, we cannot directly compare the between these two approaches because existing techniques uses the entire vessel mask for score computation, whereas we only use the vessel path(or center-line path). Additionally, to reciprocate with our original idea to be able to compute all vessel abnormalities, we extract the entirety of the vascular structure as opposed to using fixed threshold to generate a binary mask, used by the previous techniques. By choosing a specific threshold, we can easily bias our result towards one of Sensitivity or Specificity. For example, a lower threshold only segments high confidence larger vessels, making it easier for the network to classify A-V. So we believe its unfair to directly compare the results. Nonetheless, our results being more constrained, are comparable or even better than the existing techniques. The rows labeled \textbf{CNN} represent the result fro direct U-Net, whereas \textbf{CNN + Prop} represents after we do some basic label propagation mentioned in the original topology extraction paper\cite{top_https://doi.org/10.48550/arxiv.2202.02382}. We are very close in DRIVE dataset in BACC, whereas better in Sensitivity. Similarly, for INSPIRE-AV dataset, we have the state of the art result with 0.924 BACC vs 0.91 by Ma et al. \cite{multi_task_https://doi.org/10.48550/arxiv.2007.09337}. We can see similar comparable results in LES dataset as well.

\subsection{Artery-Vein ratio}
\begin{figure*}[!ht]
    
    \centering
    \includegraphics[width=0.99\textwidth]{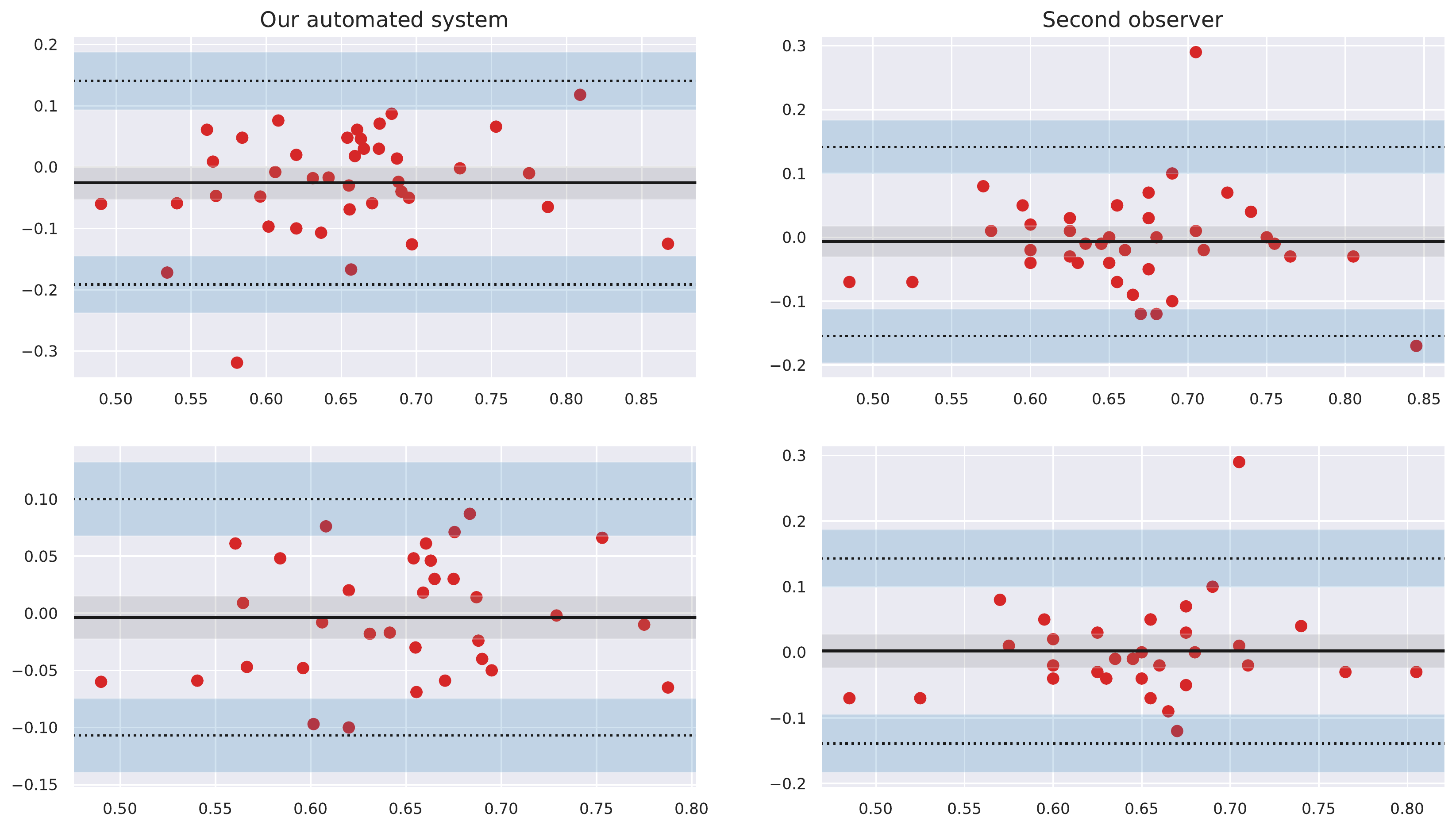}
    \caption{\textbf{Bland-Altman plot} shows the agreement between automated system, and the referenced values(left), and agreement between the referenced observer and the second observer. The dotted line shows $95\%$ limits of agreement. Top row is the comparison on the full dataset, whereas bottom one is the images with AVR error $< 0.1$ (33 out of 40 images), and 20 images error $< 0.05$.}
   \label{fig:Sys-Ref}
\end{figure*}

We have also calculated Artery-Vein ratio in INSPIRE-AVR dataset\cite{inspire_av_dataset, inspire_dataset-tang2011robust}. It was the only dataset with ground-truth available. We calculated AVR for all images and compared against two human graders, as shown in the table \ref{tab:inspire_AVR}. We obtained the average error of $0.65\pm0.058$ with the referenced grader, whereas $0.059\pm0.059$ on the second observer. Compared to the only existing work on the same dataset by  Niemeijer et al. \cite{av_ratio_5876322}, we have achieved same statistics, but with our approach we have gotten 33 out of 40 images with less than $0.01$ error with the reference observer, which was 32 in the before-mentioned work. The Bland Altman plot in figure \ref{fig:Sys-Ref} compares the result of our automated system with a reference manual labels , and a second observer. We can see the mean difference between the AVR between the referenced and second observer are close to 0, and almost 0 for the ones with $error < 0.1$ in second row, figure \ref{fig:Sys-Ref} indicating no substantial bias in automated system's result, and that of the second observer. We can also see a comparable inclusion within $95\%$ confidence range(dotted lines) between the automated system, and the second observer. 

\subsection{Vessel tortuosity} 
\begin{figure*}[!ht]
    \centering
    \includegraphics[width=0.99\textwidth]{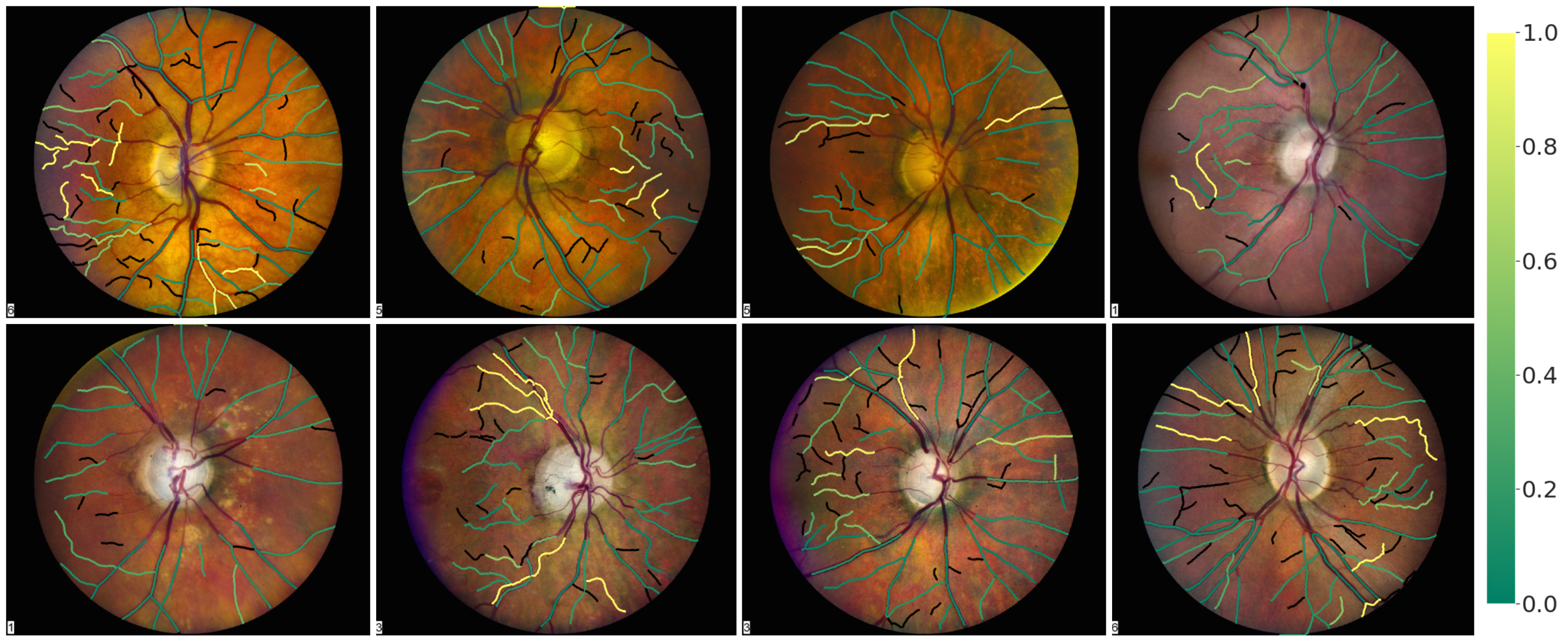}
    \caption{Tortuosity in INSPIRE-AVR dataset}
    \label{fig:Tortuosity}
\end{figure*}

We have only found a handful of paper that worked in calculating retinal vessel tortuosity. Ramos et al. presented a work on tortuosity measurement but the dataset was not publicly \cite{comp_assesment_tort_Ramos2019}. Thus, having no any reference to compare our results against, we simply have plotted the caliber of tortuoisty in the fundus image so that we can visually asses it(figure \ref{fig:Tortuosity}). The tortuosity is scaled from $0-1$ with $0$ being a non-torturous, and $1$ being highly tortourus vessel.

\begin{table}[!ht]
    \caption{AVR result in INSPIRE result}
    \label{tab:inspire_AVR}
    \centering
    \begin{tabular}{llllll}
        \toprule
        \textbf{File} & \textbf{Reference} & \textbf{System} & \textbf{Error} & \textbf{Obs. 2} & \textbf{Error2} \\ \midrule
        
        image1.jpg & 0.7 & 0.641 & 0.058 & 0.71 & 0.068 \\ \midrule
        image2.jpg & 0.63 & 0.691 & 0.061 & 0.68 & 0.011 \\ \midrule
        image3.jpg & 0.7 & 0.676 & 0.023 & 0.65 & 0.026 \\ \midrule
        image4.jpg & 0.65 & 0.633 & 0.016 & 0.64 & 0.006 \\ \midrule
        image5.jpg & 0.78 & 0.77 & 0.009 & 0.75 & 0.02 \\ \midrule
        image6.jpg & 0.65 & 0.553 & 0.096 & 0.65 & 0.096 \\ \midrule
        image7.jpg & 0.67 & 0.57 & 0.099 & 0.65 & 0.079 \\ \midrule
        image8.jpg & 0.64 & 0.686 & 0.046 & 0.71 & 0.023 \\ \midrule
        image9.jpg & 0.69 & 0.583 & 0.106 & 0.76 & 0.176 \\ \midrule
        image10.jpg & 0.56 & 0.608 & 0.048 & 0.85 & 0.241 \\ \midrule
        image11.jpg & 0.64 & 0.727 & 0.087 & 0.74 & 0.012 \\ \midrule
        image12.jpg & 0.76 & 0.634 & 0.125 & 0.75 & 0.115 \\ \midrule
        image13.jpg & 0.57 & 0.646 & 0.076 & 0.62 & 0.026 \\ \midrule
        image14.jpg & 0.62 & 0.572 & 0.047 & 0.58 & 0.007 \\ \midrule
        image15.jpg & 0.64 & 0.622 & 0.017 & 0.61 & 0.012 \\ \midrule
        image16.jpg & 0.68 & 0.694 & 0.014 & 0.68 & 0.014 \\ \midrule
        image17.jpg & 0.52 & 0.46 & 0.059 & 0.45 & 0.01 \\ \midrule
        image18.jpg & 0.62 & 0.448 & 0.171 & 0.63 & 0.181 \\ \midrule
        image19.jpg & 0.67 & 0.64 & 0.029 & 0.63 & 0.01 \\ \midrule
        image20.jpg & 0.71 & 0.67 & 0.039 & 0.62 & 0.05 \\ \midrule
        image21.jpg & 0.57 & 0.511 & 0.058 & 0.58 & 0.068 \\ \midrule
        image22.jpg & 0.72 & 0.67 & 0.049 & 0.76 & 0.089 \\ \midrule
        image23.jpg & 0.66 & 0.69 & 0.03 & 0.69 & 0.0 \\ \midrule
        image24.jpg & 0.65 & 0.68 & 0.03 & 0.64 & 0.04 \\ \midrule
        image25.jpg & 0.56 & 0.569 & 0.009 & 0.49 & 0.079 \\ \midrule
        image26.jpg & 0.73 & 0.728 & 0.001 & 0.61 & 0.118 \\ \midrule
        image27.jpg & 0.64 & 0.711 & 0.071 & 0.63 & 0.081 \\ \midrule
        image28.jpg & 0.63 & 0.678 & 0.048 & 0.68 & 0.001 \\ \midrule
        image29.jpg & 0.72 & 0.786 & 0.066 & 0.7 & 0.086 \\ \midrule
        image30.jpg & 0.59 & 0.543 & 0.046 & 0.61 & 0.066 \\ \midrule
        image31.jpg & 0.75 & 0.868 & 0.118 & 0.75 & 0.118 \\ \midrule
        image32.jpg & 0.53 & 0.591 & 0.061 & 0.61 & 0.018 \\ \midrule
        image33.jpg & 0.61 & 0.602 & 0.007 & 0.59 & 0.012 \\ \midrule
        image34.jpg & 0.65 & 0.668 & 0.018 & 0.61 & 0.058 \\ \midrule
        image35.jpg & 0.74 & 0.573 & 0.166 & 0.64 & 0.066 \\ \midrule
        image36.jpg & 0.69 & 0.621 & 0.068 & 0.62 & 0.001 \\ \midrule
        image37.jpg & 0.82 & 0.755 & 0.064 & 0.79 & 0.034 \\ \midrule
        image38.jpg & 0.93 & 0.805 & 0.124 & 0.76 & 0.045 \\ \midrule
        image39.jpg & 0.61 & 0.63 & 0.02 & 0.64 & 0.009 \\ \midrule
        image40.jpg & 0.74 & 0.421 & 0.318 & 0.62 & 0.198 \\ \midrule
        \textbf{Mean} & 0.666 & 0.641 & 0.065 & 0.66 & 0.059 \\
        \textbf{STD} & 0.079 & 0.093 & 0.058 & 0.077 & 0.059 \\
        \textbf{Min} & 0.079 & 0.093 & 0.001 & 0.077 & 0.0 \\
        \textbf{Max} & 0.93 & 0.868 & 0.318 & 0.85 & 0.241 \\
        \bottomrule
    \end{tabular}
\end{table}

\section{Discussion}
Most of the research on retinal image analysis focuses on vessel segmentation, Artery-Vein classification, optic disk segmentation, and topology estimation. There are very few research works that solely focus on features like AV ratio, vessel tortuosity. Those ones that did have either private dataset used \cite{comp_assesment_tort_Ramos2019}, or patches on parts of the vessels\cite{vessel_seg_tort_app10144788}. Our work, however, proposes a complete vascular extraction and feature calculation framework. Moreover, we have shown that a completely extracted topology is a better input for Artery-Vein classification in table \ref{tab:literature-results}. The results yielded are Artery or Vein labels assigned to each nodes of the topology, which can be a very strong prior for other down stream tasks like label propagation, vessel traversal and measurements as shown in the results above.

\section{Conclusion and Future work}
We have advanced the research on retinal vessel analysis in the correct direction by extracting the most useful features--Artery-Vein ratio, and vascular tortuosity. The topology extracted by our technique is benefited by AV classification, AV-ratio calculation, and vessel tortuosity measurement as shown in this paper. We strongly believe that we can extract other vasculature based features from the same topology. There are various path we can take beyond this; We can use these features to computationally asses the pathologies in the eye. We can also extract further more features and do similar analysis.

\bibliographystyle{IEEEtran}
\bibliography{report}
\end{document}